\documentclass[preprint]{elsarticle}
\usepackage{graphicx}%  
\usepackage[pdfa]{hyperref}%  
\usepackage{xcolor}

\newcommand{\vol}[3]{\textbf{#1}\textrm{ (#3) #2}}
\newcommand{\EPJ} {Eur. Phys. J.}

\newcommand{\NP} {Nucl. Phys.}
\newcommand{\PL} {Phys. Lett.}
\newcommand{\PR} {Phys. Rev.}

\newcommand{\PRL} {Phys. Rev. Lett.}

\newcommand{\XeSn} {$^{129}$Xe+$^{\rm nat}$Sn }
\newcommand{\AuAu} {$^{197}$Au+$^{197}$Au }
\newcommand{\Et} {$E_{\perp 12}$}
\newcommand{\Mc} {$M_{c}$ }
\newcommand{\bbmax} {$b/b_{\rm max}$ }

\begin{document}
\begin{frontmatter}

\title{Isotopic Transparency in Central Xe+Sn Collisions at 100 MeV/nucleon}
%at Intermediate Energy}

\author[a]{A.~Le~F\`{e}vre\corref{corr}}
\author[b]{A.~Chbihi}
\author[c]{Q.~Fable}
\author[b]{T.~G\'{e}nard}
%\author[b]{O.~Lopez}
\author[d]{J.~{\L}ukasik}
\author[a]{W.~Trautmann}
\author[a]{K.~Turz\'{o}}
\author[e]{R.~Bougault}
\author[f]{S.~Hudan}
\author[e]{O.~Lopez}
\author[a]{W.F.J.~M\"{u}ller}
\author[a]{C.~Schwarz}
\author[g]{C.~Sfienti}
\author[c,h]{G.~Verde}
\author[i]{M.~Vigilante}
\author[j]{B.~Zwiegli\'{n}ski}

\cortext[corr]{corresponding author: A.LeFevre@gsi.de}

%The INDRA and ALADIN Collaborations

\address[a]{GSI Helmholtzzentrum f\"{u}r Schwerionenforschung GmbH, D-64291 Darmstadt, Germany}
\address[b]{GANIL, CEA et IN2P3-CNRS, F-14076 Caen, France}
\address[c]{Laboratoire des 2 Infinis - Toulouse (L2IT-IN2P3), Universit\'e de Toulouse, CNRS, UPS, F-31062 Toulouse Cedex 9, France}
\address[d]{H. Niewodnicza\'{n}ski Institute of Nuclear Physics, Pl-31342 Krak\'{o}w, Poland}
\address[e]{Universit\'{e} de Caen Normandie, ENSICAEN, CNRS/IN2P3, LPC Caen UMR6534, F-14000 Caen, France}
\address[f]{Department of Chemistry and Center for Exploration of Energy and Matter, Indiana University, Bloomington, Indiana 47408, USA}
\address[g]{Institute of Nuclear Physics, Johannes Gutenberg University, D-55099 Mainz, Germany}
\address[h]{INFN Catania, Italy}
\address[i]{Dipartimento di Scienze Fisiche e Sezione INFN, Univ. Federico II, I-80126 Napoli, Italy}
\address[j]{National Centre for Nuclear Research, PL-02093 Warsaw, Poland}

\date{}
%\maketitle

\begin{abstract} 

A new method, based on comparing isotopic yield ratios measured at forward and sideward polar angles and 
on cross-bombarding heavy nuclei with different neutron-to-proton ratios, is used to quantify
the stopping power of nuclear matter in heavy-ion collisions. For central collisions of
isotopically separated $^{124,129}$Xe+$^{112,124}$Sn at 100~MeV/nucleon 
bombarding energy, measured with the 4$\pi$ multidetector INDRA at GSI, a moderate 
transparency is deduced for hydrogen isotopes, whereas for heavier fragmentation products
with atomic number $Z \ge 3$ a high transparency exceeding 50\% is observed.
An anomalously large transparency is found for alpha particles, and possible
explanations are presented. 
 
\end{abstract}

\begin{keyword}
NUCLEAR REACTIONS \sep $^{112,124}$Sn($^{124,129}$Xe, X), $E=100$~MeV/nucleon
\sep multifragmentation \sep breakup state \sep isotopic effects
\sep transparency \sep nuclear matter stopping power  
\end{keyword}

\end{frontmatter}

%\noindent
%{\color{red} {\bf Introduction}}

%\noindent
%{\color{red} paragraph on stopping and its importance}

Stopping is an important property identifying the nature of heavy-ion collisions~\cite{likobauer98}. 
Its variation with the incident energy characterizes the evolution of the collision dynamics.
At incident energies above the Coulomb barrier, the dependence of compound nucleus formation and dissipative 
binary collisions on impact parameter 
has been described within the concepts of critical angular momentum 
and surface friction~\cite{wilczynski,volkov78,gross_kalinowski,siwek76,trautmann93}. 
%and surface friction~\cite{wilczynski,gross_kalinowski,siwek76,trautmann93}. 
Above the Fermi-energy domain, nucleon-nucleon collisions
govern the dynamics, and stopping in central collisions is of particular interest because dense 
matter may be formed in the collision zone, offering
the possibility of studying the equation of state of nuclear matter beyond its saturation properties 
(Refs.~\cite{baoanli02,dani02,le_fevre16,huth22} and references therein).

%\noindent
%{\color{red} usual definition of stopping and survey of existing results}

An observable conveniently used to measure the degree of relaxation of the initial momentum is the isotropy ratio 

\begin{equation}
      R_p = \frac{<p_x^2>+<p_y^2>}{2<p_z^2>} 
\label{eq:eqR}
\end{equation}

\noindent describing the strengths of the variances of transverse momenta $p_x$ and $p_y$ relative to that 
of the longitudinal momenta $p_z$ of reaction products in the center-of-mass (c.m.)
frame~\cite{reisdorf04,andronic06,zbiri}. The ratio $R_p = 1$ indicates complete stopping with the production 
of an isotropic momentum distribution, a situation hardly reached at intermediate bombarding energies. 
The excitation function for stopping in central \AuAu collisions over the incident energy range from 15 MeV/nucleon 
up to 1.5 GeV/nucleon exhibits a rise from $R_p \approx 0.5$ near 40 MeV/nucleon to a
maximum of 0.9 at 600 MeV/nucleon and a wide plateau with values exceeding 0.8 
at energies between 200 MeV/nucleon and 1 GeV/nucleon (FOPI and INDRA data, Andronic et al.~\cite{andronic06}).
The energy range up to 100 MeV/nucleon was investigated in detail by Lehaut et al.~\cite{lehaut10} 
who used data for several mass-symmetric reactions from Ar+KCl to \AuAu which had been studied with the INDRA 
multidetector~\cite{pouthas}. 
A general trend of isotropy ratios near 0.6 exhibiting a weak rise with incident energy is reported for the selected central event classes. 

%\noindent
%{\color{red} conclusion on elongated sources}

The formation of composite systems with longitudinally elongated momentum distributions is thus found
to be a general feature of central heavy-ion collisions at intermediate energies. 
It can be the result of a transparency, 
i.e. of an incomplete stopping of the incoming projectile by the target 
as, e.g., observed by the FOPI collaboration for Zr+Ru central collisions at 
400~MeV/nucleon~\cite{rami}.
This phenomenon is manifested by an incomplete relaxation of momenta and 
by angular anisotropies of fragment yields, as reported by the INDRA 
collaboration for \AuAu and \XeSn central collisions at energies between 40 and 
150~MeV/nucleon~\cite{zbiri,le_fevre04,le_fevre07,lavaud,bouriquet}.

%\noindent
%{\color{red} report on studies explaining the prevalence of elongated sources}

Several approaches were used to interpret the anisotropies observed in 
fragment productions. The study of central \XeSn ~and \AuAu ~collisions at bombarding 
energies between 50 and 100 MeV/nucleon demonstrated that an accurate 
statistical description of the measured anisotropies in fragment yields and kinetic 
energies is obtained if a prolate source deformation and a superimposed collective motion 
are included~\cite{le_fevre04}. The element yields were found to extend to larger atomic numbers 
$Z$ at forward and backward emission angles than at sideward angles. 
In the Metropolis Multifragmentation Monte Carlo (MMMC) 
model used 
in that work, an important role is played by the Coulomb interaction which favors large spatial
separations between heavy fragments in order to minimize the Coulomb energy~\cite{le_fevre04}. 
Heavy fragments are preferentially placed in the tips of a prolate source. 
Through the Coulomb repulsion and the superimposed radial flow, these spatial correlations 
induce correlations in momentum space
which lead to the observed maxima in the yields and kinetic energies of the
heaviest fragments at forward and backward directions. 

This scenario was confirmed in a study using correlation functions constructed from directional projections 
of the relative velocities of fragments which was applied to central \XeSn collisions at 50~MeV/nucleon~\cite{le_fevre07}. 
It showed that the predicted correlation functions obtained from MMMC calculations depend strongly on the chosen 
source geometry at breakup. Assuming a prolate deformation of the breakup volume in coordinate space, 
aligned along the beam direction, permitted a satisfactory description of the experimental data.

With a quantum molecular dynamics approach, the anisotropies observed in central \AuAu collisions between 60 
and 150~MeV/nucleon bombarding energy were successfully reproduced~\cite{zbiri}.
It was shown that the fragment compositions close to midrapidity can be described by statistical laws whereas 
they are dominated by the dynamics at large and small rapidities. A sizeable fraction of the initial 
momentum is not relaxed: the majority of fragments represent surviving initial-final state correlations and are not 
substantially decelerated in longitudinal direction. In this dynamical approach, the stopping power of nuclear matter is found to be strong 
for hydrogen isotopes but decreasing with atomic number $Z$ and very small for heavier fragments.   

%\noindent
%{\color{red} aim of the present study to apply it quantitatively to XeSn with an complete data set for the four combinations}

\begin{table}[!ht]
  \caption{For each system, from left to right: $N/Z$ value of the combined system, thresholds of \Mc
    and of \Et~corresponding to a reduced impact parameter \bbmax~= 0.1 (see text). 
    The enrichment of the Sn targets is taken into account. 
  }
  \begin{center}
    \begin{tabular} {l c c c}
      \hline\hline
      \\
      System & $N/Z$ & \Mc & \Et~(MeV)  \\
      \hline
      \\
      $^{124}$Xe+$^{112}$Sn & 1.270 & 50 & 1463  \\
      $^{129}$Xe+$^{112}$Sn & 1.318 & 50 & 1455  \\
      $^{124}$Xe+$^{124}$Sn & 1.385 & 50 & 1465  \\
      $^{129}$Xe+$^{124}$Sn & 1.433 & 49 & 1451  \\
      \\
      \hline\hline
    \end{tabular}
  \end{center}
  \label{tab:table1}
\end{table}

The aim of the present work is to quantify the stopping power of nuclear matter in central collisions,
on a pure experimental basis, 
by using the isotopic composition of the emitted particles and fragments as observables
capable of identifying the primary sources in which they are formed~\cite{li_yennello95,johnston96,ogul23}.
The degree of mixing of nucleons of the target and of the projectile that has occurred in the course of the collision
and the observed chemical equilibration will complement the information provided by the momentum observables.
Measurements at sideward angles, at which full mixing follows from the approximate mass symmetry of the four collision systems,
are used as a base line for the interpretation. 
A statistical description is found useful
for this purpose but we do not intend to explain in this work the mechanism by which clusters are formed.

%\noindent
%{\color{red} {\bf Experimental method}}

We report data for Xe+Sn collisions at 100~MeV/nucleon incident energy, 
obtained at the GSI laboratory with the INDRA multidetector~\cite{pouthas}. By cross-bombarding 
enriched targets of $^{112}$Sn (98.9\%) and $^{124}$Sn (99.9\%) with $^{124,129}$Xe projectiles, 
neutron-over-proton ratios $N/Z$ of the combined system ranging from
1.27 to 1.43 were covered (Table~\ref{tab:table1}). 
Details of the setup and data calibration 
for this experiment are found in Ref.~\cite{le_fevre04}. 

%\noindent
%{\color{red} selection of central collisions}

Central collisions were selected either according to the total multiplicity of 
charged particles \Mc or with the total transverse kinetic energy $E_{\perp 12}$ of light charged particles with
$Z=1,2$. In both cases, only 1\% of the recorded reaction cross sections were accepted,
corresponding to a reduced impact parameter \bbmax ~$\leq 0.1$ in sharp-cut-off approximation and assuming
a monotonic relation of \Mc or $E_{\perp 12}$ with the impact parameter. The acceptance thresholds were found to
be rather similar for the four studied reactions with thresholds \Mc $\approx 50$ and $E_{\perp 12} \approx 1460$~MeV
(Table~\ref{tab:table1}).
It is obvious that these high thresholds 
cut into the tails of the \Mc and \Et~distributions. As demonstrated in \cite{zbiri}, 
these tails may be partly caused by the fluctuation widths of \Mc and \Et~in central collisions rather than by a significant variation 
of the impact parameter up to the highest values of these variables. 

\begin{figure}[!htb]	%Fig. 1
  \leavevmode
  \begin{center}
    \includegraphics[width=11cm]{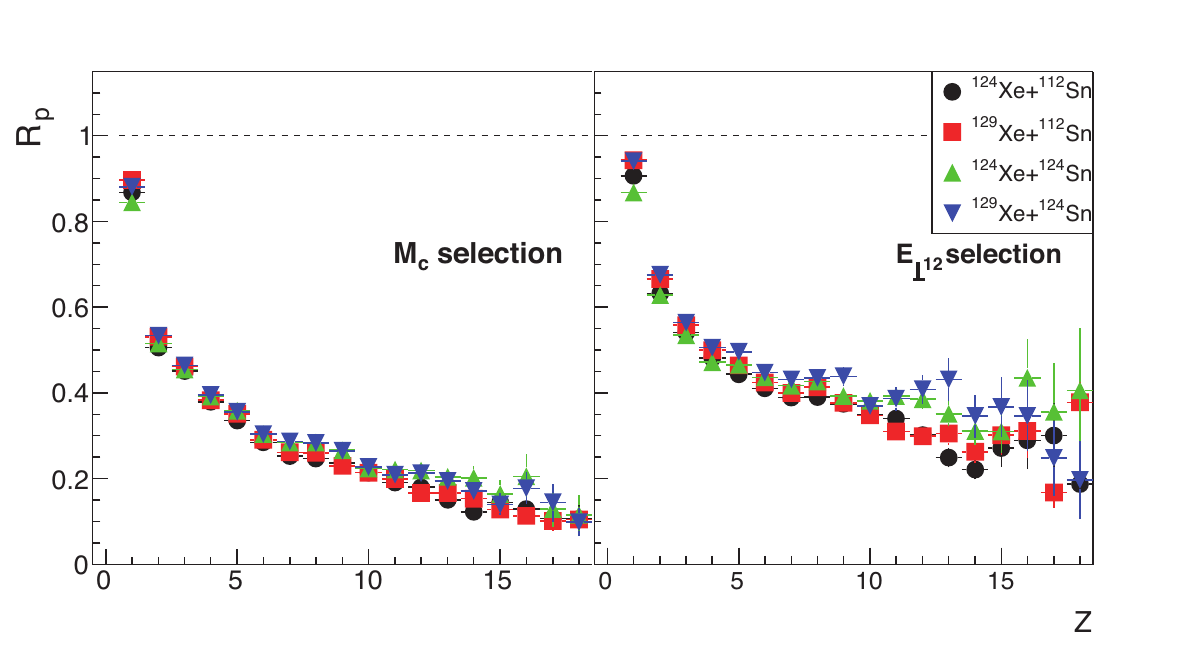} % Fig1mod.eps for WT
  \end{center}
  \caption{Isotropy ratio $R_p$ 
    as a function of the fragment atomic number $Z$ for the four combinations 
    of $^{124,129}$Xe+$^{112,124}$Sn reactions at 100~MeV/nucleon incident energy 
    in central collisions selected according to \Mc (left panel)
    and \Et~(right panel). 
  }
  \label{fig:fig1} 
\end{figure}  

The isotropy ratio $R_p$ defined in Eq.~(\ref{eq:eqR}) 
is displayed in Fig.~\ref{fig:fig1} as a function of the fragment $Z$ 
for the central event classes of the four systems, selected according to \Mc or \Et. 
In the latter case, 
in order to avoid autocorrelations between the $R_p$ values for $Z=1,2$ and $E_{\perp 12}$, 
the centrality criterion 
$E_{\perp 12}$ was calculated without the particle of interest.
A $Z=1$ or $Z=2$ particle whose exclusion from $E_{\perp 12}$ makes the event no longer ``central'' is not taken into account 
in the calculation of $R_p$ for this particular $Z$.

%\noindent
%{\color{red} effect on $R_p$ (Fig. 1)}

We observe the same trend for both selections and all systems. 
The isotropy ratio $R_p$ is close to 1 for hydrogen isotopes, indicating full relaxation of the initial momenta, 
but decreases rapidly with the fragment atomic number. The momenta of the larger fragments
are predominantly oriented in longitudinal directions.  
The ratio $R_p$ is found to be significantly higher with \Et~than with \Mc centrality selections. 
This is possible because the \Mc and \Et~selected event groups
are only partly overlapping as documented for similar reactions in Fig.~3 of Ref.~\cite{le_fevre04}. The transfer of
transverse momentum to light particles with $Z=1,2$ and to fragments with intermediate mass is
apparently correlated. Choosing reactions with highest \Et~thus favors the cases for which the largest stopping has occurred.
In addition, there is no visible influence of the initial isotopic compositions on $R_p$ in the case of the \Mc selection
whereas, with the \Et~selection, the ratio $R_p$ increases with the $N/Z$ of the collision system for the larger $Z$ 
values (Fig~\ref{fig:fig1}). The mode of selecting central events has a measurable
influence on the resulting momentum configuration. 

\begin{figure}[!htb]	%Fig. 2
    \leavevmode
\begin{center}
\includegraphics[width=8.0cm]{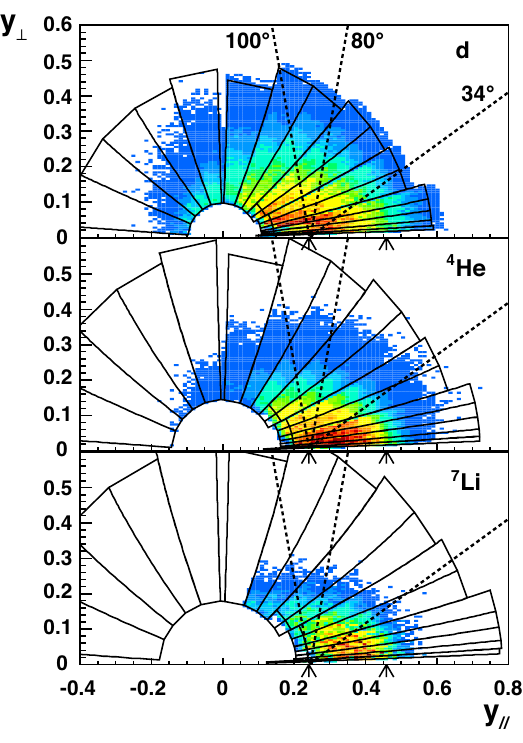} % Fig2a.eps for WT
\end{center}
\caption{Two-dimensional invariant distributions in the transverse versus longitudinal rapidity space of deuterons, 
$^4$He, and $^7$Li isotopes (top to bottom) for Xe+Sn central collisions at 100~MeV/nucleon incident energy.
Dashed lines represent the forward and sideward acceptance cuts in the center-of-mass frame
($\theta_{\rm cm}\leq34^{\circ}$ and $80^{\circ}\leq\theta_{\rm cm}\leq100^{\circ}$, respectively).
Full lines depict the geometrical acceptance of the 17 rings 
of the INDRA multidetector, including high and low energy thresholds.
The arrows below the x-axes at $y_{\|} = 0.46$ and 0.23 mark the longitudinal rapidities of the projectiles and
of the center-of-mass system assuming mass symmetry, respectively. 
}
\label{fig:fig2} 
\end{figure}  

For the elements with atomic number $Z \le 4$ considered in the following, the INDRA multidetector offers excellent 
isotopic resolution~\cite{pouthas}. 
In the nine forward rings (at polar angles $\theta_{\rm lab} \le 45^{\circ}$), the mass 
identification for light elements with sufficiently high velocity is possible using the maps 
of the fast versus the slow components of the light detected in the detectors made of cesium-iodide scintillating crystals doped with thallium (CsI(Tl)). 
Alternatively, the maps of the signals recorded by the 300-$\mu$m silicon detectors versus the total 
light in the CsI(Tl) detectors may be used. At angles beyond $45^{o}$, at which INDRA is not equipped with silicon detectors, the fast versus slow maps of the CsI(Tl) signals are used for mass identification.

%\noindent
%{\color{red} discuss acceptance (Fig. 2)}

The measured two-dimensional invariant distributions in the transverse versus longitudinal rapidity space 
of deuterons, $^4$He, and $^7$Li isotopes 
are displayed in Fig.~\ref{fig:fig2}.
The yields of the three isotopes are well covered by the acceptance of the INDRA multidetector
within both, the forward and sideward angular cuts in the center-of-mass system chosen for the analysis: 
polar angles $\theta_{\rm cm} \leq 34^{\circ}$
and $80^{\circ}\leq\theta_{\rm cm}\leq100^{\circ}$, respectively.
As shown in Fig.~\ref{fig:fig2}, center-of-mass emissions to sideward angles are spread over most of the rings in the forward hemisphere of the INDRA detector and  identified on the basis of the measured kinetic energies. The c.m. yields in forward direction are spread over fewer INDRA rings and selected with a ring dependent lower threshold of the kinetic energies. The angular cuts are clearly 
approximate within these limits but essentially identical for the four reactions.

The coverage is similar for the remaining isotopes up to $^{7,10}$Be.
The decreasing degree of stopping with increasing fragment mass is as well evident in Fig.~\ref{fig:fig2}. In fact, 
the sideward emissions of, e.g., $^{7}$Li
fragments are no longer representing the central part of the distribution but originate from the tail 
connecting the forward and backward moving groups. 

%\noindent
%{\color{red} {\bf Experimental results}}

%\noindent
%{\color{red} observation at forward and sideward angles, show result (Fig. 3) and discuss slopes (Fig. 4)}

\begin{figure}[!t]	%Fig. 3
  \begin{center}
    \includegraphics[width=9.0cm]{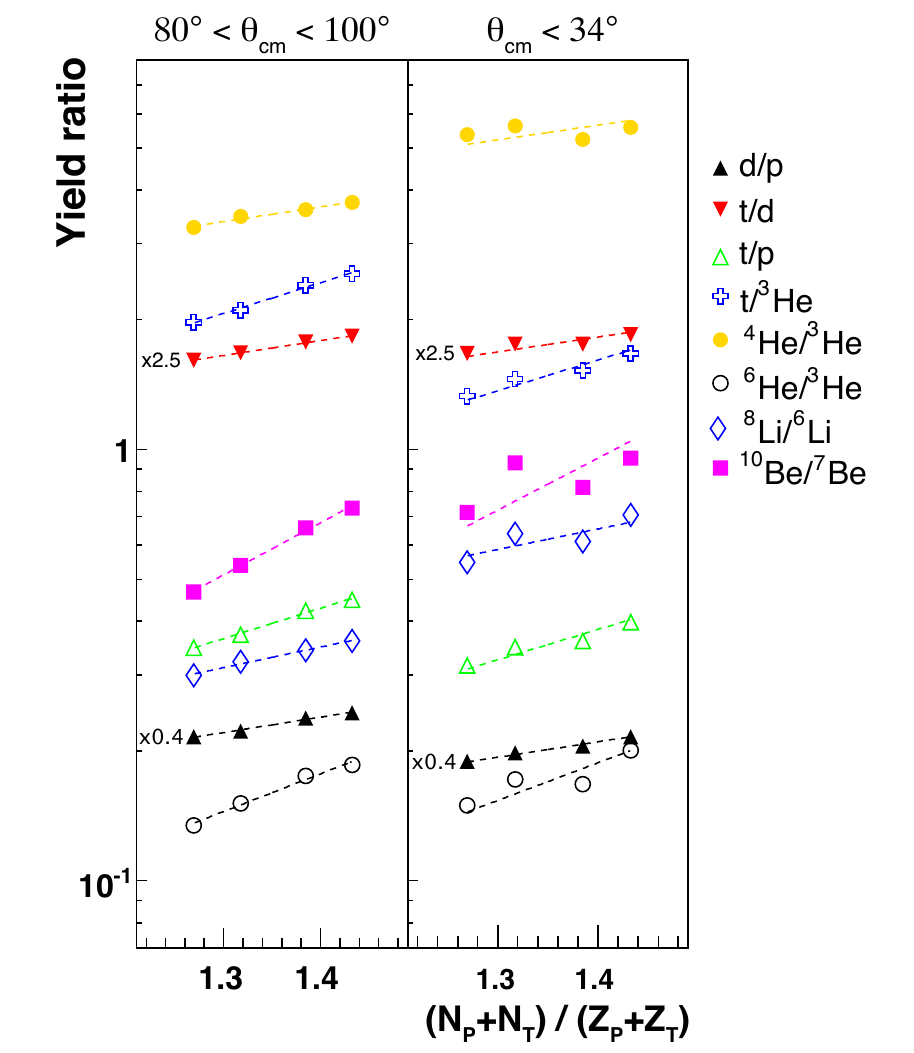} % Fig3a.eps for WT
    \caption{Isotope yield ratios as functions of the total $N/Z$ of the system 
      in central collisions, selected according to $M_c$, of $^{124,129}$Xe+$^{112,124}$Sn at 100~MeV/nucleon, 
      for sideward angles (left panel, $80^{\circ}\leq\theta_{\rm cm}\leq100^{\circ}$) 
      and forward angles (right panel, $\theta_{\rm cm}\leq 34^{\circ}$). 
      To avoid overlaps of data symbols, the indicated factors were applied to the measured d/p and t/d ratios.
      Statistical errors range from below 1\% for light particles up to $\approx 3$ or 4\% for the Li 
      and Be ratios and are smaller than the size of the symbols. 
      The lines represent exponential fits to the ratios. In the right panel, to facilitate comparisons, 
      fits are constrained to reproduce the slopes found in the left panel.
      } 
    \label{fig:fig3}
  \end{center}
\end{figure}

We present in Fig.~\ref{fig:fig3}, left panel, the yield ratios
of several light isotopes emitted at sideward angles 
in the center-of-mass ($80^{\circ}\leq\theta_{\rm cm}\leq100^{\circ}$) 
as a function of the total $N/Z$ of the system 
for central collisions selected with $M_c$ (cf. Table~\ref{tab:table1}).
Their exponential dependence on $N/Z$, observed for all cases, is remarkable. It reflects 
the full mixing of target and projectile matter expected at mid-rapidity~\cite{wada87,trautmann01}.

\begin{figure} 		%Fig. 4
 \begin{center}
     \includegraphics[width=7.0cm]{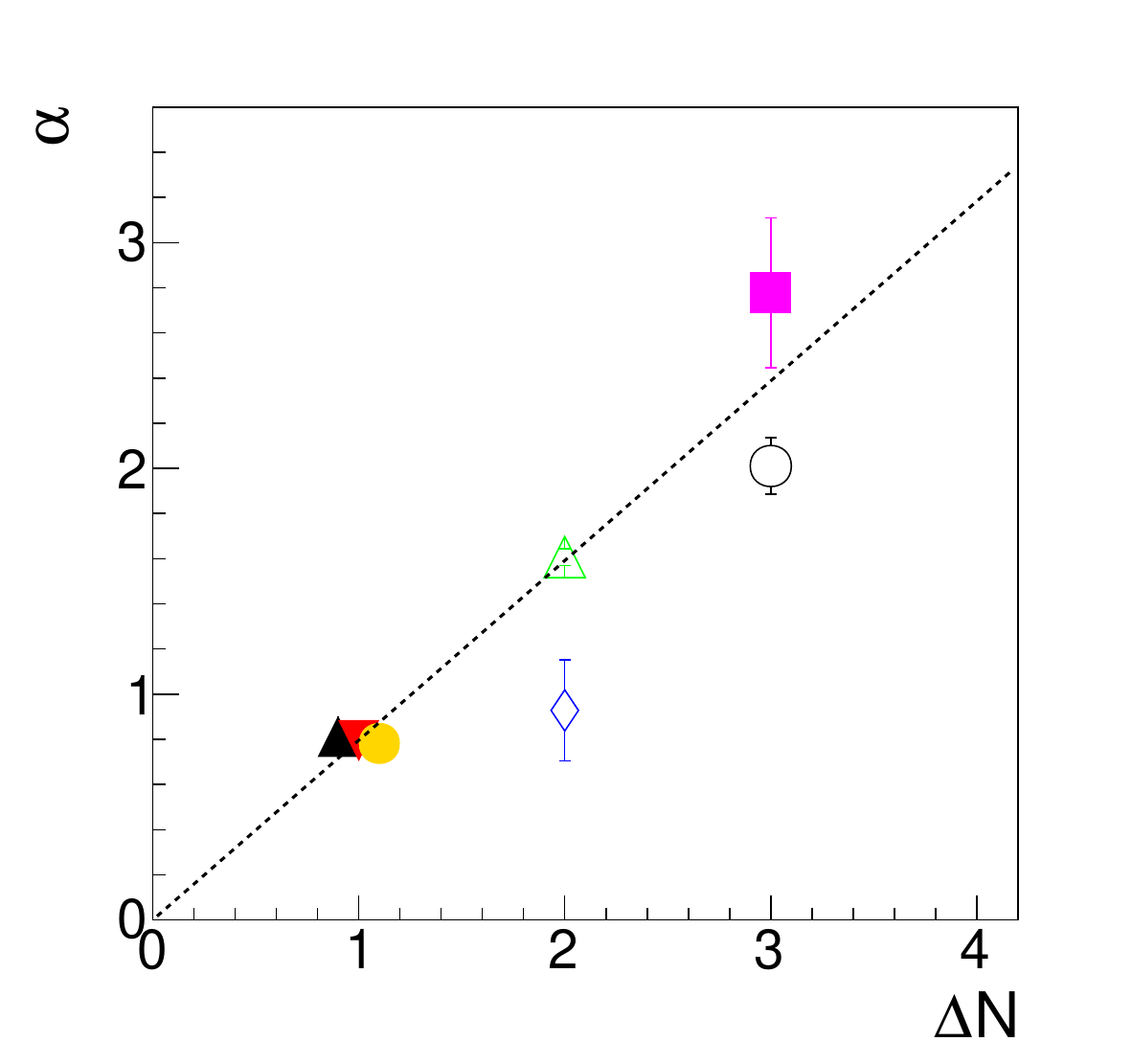} % Fig4.eps for WT 
     \caption{Slopes $\alpha$ of exponential fits extracted from Fig.~\ref{fig:fig3}, left panel (sideward emissions,
       $M_c$ selection), as a function of the difference $\Delta N$ in neutron number of the isotope pair. 
       The t/$^3$He isobar ratio is not included.
       The displayed errors are those returned by the fit routine.
       The dotted line represents the weighted mean of the rise of $\alpha$ with $\Delta N$.
       The symbols are chosen as in Fig.~\ref{fig:fig3}. For better visibility,
       the symbols related to the t/d and $^4$He/$^3$He ratios are slightly shifted horizontally. 
       } 
     \label{fig:fig4}
 \end{center}
\end{figure}

%\noindent
%{\color{red} systematic error derived from choosing different centrality selection schemes}

The logarithmic slope parameters $\alpha$ of exponential fits are found to be proportional
to the difference $\Delta N$ of the neutron numbers of the two isotopes forming a pair (Fig.~\ref{fig:fig4}), 
suggesting a statistical emission pattern in the grand-canonical description~\cite{barz88}.
Qualitatively the same results are obtained when $E_{\perp 12}$ is used to select central events, with the exception
that the measured ratios are, on average, 4\% larger with $E_{\perp 12}$ than with $M_c$ as selection criterion.
The average value of the absolute differences between the ratios obtained with the two selections is $\approx 6\%$. 
This is larger than the statistical errors and may thus serve as an estimate of the overall experimental uncertainties.

%\noindent
%{\color{red} show grand-canonical predictions (Table 2)}

The application of grand-canonical sampling permits a global understanding of the measured isotope ratios.
In its simplest form, the yield ratios of isotopes differing by $\Delta N$ neutrons can be expressed as

\begin{equation}
\frac{Y(A_2,Z)}{Y(A_1,Z)} = (\frac{A_2}{A_1})^{3/2}~\frac{\omega (A_2,Z)}
{\omega (A_1,Z)}~\exp(\frac{\Delta N\times\mu_{\rm n} + \Delta B}{T})
\label{eq:grandcan}
\end{equation}

\noindent with $A_1$ and $A_2 = A_1 + \Delta N$ representing the mass numbers of the two isotopes, 
$\omega (A_i,Z)$ their internal partition functions, and $\Delta B = B_2 - B_1$
the difference of their binding energies~\cite{mekjian78,albergo}.
The chemical potentials $\mu_{\rm n}$ of neutrons and $\mu_{\rm p}$ of protons
guarantee the conservation of the mean mass 
and charge of the disassembling system within a given volume and thus are functions of its $N/Z$ ratio. 
The chemical potential $\mu_{\rm p}$ of protons
appears only in the t/$^3$He isobar ratio involving nuclides with different $Z$.
%, showing similar features as in~\cite{xu00}.
The assumption that the chemical potentials depend, to first order, linearly on $N/Z$ immediately
explains the experimental finding that, in the apparently equilibrated case at sideward angles,
the logarithm of the ratio increases linearly with the 
$N/Z$ ratios of the emitting sources.

The grand-canonical predictions shown in Table~\ref{tab:table2} were obtained in the following way.
For the temperature $T$, the double-isotope temperature~\cite{albergo,kunde98,trautmann07} 

\begin{equation}
T_{\rm Hedt} = 14.31~{\rm MeV}/\ln(1.59~\frac{Y_{\rm d}/Y_{\rm t}}
{Y_{^{3}{\rm He}}/Y_{^{4}{\rm He}}})
\label{eq:hedt}
\end{equation}

\noindent was chosen and determined from the
experimental values at $N/Z = 1.35$ as given by the exponential fits of the four measured ratios (Fig.~\ref{fig:fig3}, left panel).
A value $T_{\rm Hedt} = 6.82 \pm 0.28$~MeV was obtained 
and used to determine the chemical potential $\mu_n = -10.20 \pm 0.32$~MeV for neutrons
from either one of the t/d or $^4$He/$^3$He yield ratios (the errors are obtained using the $6\%$ uncertainty of the yield ratios quoted above; 
see Ref.~\cite{trautmann07} for a discussion of systematic uncertainties of double-isotope temperatures). 
These values together with the t/$^3$He yield ratio determine also 
the chemical potential $\mu_p = -14.95 \pm 0.51$~MeV 
for protons~\cite{trautmann07}.
%The \Et~selected ratios lead to the very similar values $T_{\rm Hedt} = 6.70$~MeV and $\mu_n = -9.95$~MeV.
The ground-state spin $j$ was used for the internal partition functions 
$\omega = 2j+1$ for light particles with $Z \le 2$. 
For the heavier isotopes $^{6,8}$Li and $^{7,10}$Be, all states up to
the lowest particle threshold were included with a weight exp($-E_x/T_{\rm Hedt}$) 
given by their excitation energies $E_x$. For $^{10}$Be, e.g.,
a total of six states including the ground state were considered. The same weights were applied to determine weighted averages for the 
binding energies of these four isotopes.  
Nevertheless, since secondary decays are not taken into account, the obtained results can only be approximate (cf. Ref.~\cite{barz88}). % WT
The very similar chemical potentials $\mu_n = -11.2$~MeV and $\mu_p = -16.0$~MeV for a temperature $T=6$~MeV  for sideways emissions in the same reaction were recently reported by Bonnet et al.~\cite{bonnet25}.

\begin{table}[!ht]
  \caption{Predictions according to Eq.~(\protect\ref{eq:grandcan}) for the indicated pairs of isotopes for the temperature $T = 6.82$~MeV
and the chemical potential $\mu_n = -10.20$~MeV for neutrons in comparison with the experimental yield ratios (last column; \Mc selection) 
as given by the exponential fits for $N/Z = 1.35$ (Fig.~\protect\ref{fig:fig3}, left panel). The t/$^3$He isobar ratio is used to derive
the chemical potential $\mu_p = -14.95$~MeV for protons. 
$\Delta B$ is given in MeV; $\omega_2/\omega_1$ denotes the ratio $\omega(A2,Z)/\omega(A1,Z)$ of the internal partition functions appearing in Eq.~\ref{eq:grandcan}; see text for further details.
  }
  \begin{center}
    \begin{tabular} {l c c c c c}
      \hline\hline
\\
isotopes & $(A_2/A_1)^{3/2}$ & $\omega_2/\omega_1$ & $\Delta B$ & pred. & exp. \\
\hline
\\
 d/p & 2.83 & 1.50 & 2.22 & 1.32 & 0.57 \\
 t/d & 1.84 & 0.67 & 6.26 & 0.69 & 0.69 \\
 t/p & 5.20 & 1.00 & 8.48 & 0.90 & 0.39 \\
 t/$^3$He & 1.00 & 1.00 & 0.76 & 2.24 & 2.24 \\
 $^4$He/$^3$He & 1.54 & 0.50 & 20.57 & 3.52 & 3.52 \\
 $^6$He/$^3$He & 2.83 & 0.50 & 21.55 & 0.38 & 0.16 \\
 $^8$Li/$^6$Li & 1.54 & 0.88 & 10.46 & 0.31 & 0.32 \\
 $^{10}$Be/$^7$Be & 1.71 & 0.98 & 24.72 & 0.71 & 0.59 \\
\\
      \hline\hline
    \end{tabular}
  \end{center}
  \label{tab:table2}
\end{table}

The yield ratios measured for t/d, $^4$He/$^3$He, and t/$^3$He are reproduced exactly by construction, and others are well approximated
(last two columns of Table~\ref{tab:table2}). The d/p and t/p ratios are overpredicted by factors of $\approx 2.3$. 
There seem to be more than 
twice as many protons emitted as expected. The special role of protons has already been noticed in previous 
cases~\cite{wada87,albergo} and was
shown to be related to early out-of-equilibrium emissions in Ref.~\cite{wada87}. 
Calculations with dynamical models capable of describing the fragment production in comparable reactions \cite{zbiri,le_fevre19,ono19} could be useful for addressing this issue.

The $^6$He/$^3$He yield ratio is also overpredicted by a similar factor. It is somewhat
surprising because, for the binding energy and statistical weight, 
only the ground state of $^6$He is taken into account. One may suspect that further interactions until kinetic freeze-out
reduce the yields of $^6$He which is known to have the extended spatial distribution of a halo nucleus~\cite{tanihata85a},
actually the only isotope of this kind included in this study.
It is still remarkable that the latter three ratios obey the 
observed systematics of the slopes $\alpha = \Delta N \times (0.79 \pm 0.02)$ (Fig.~\ref{fig:fig4}) which, 
in the grand-canonical approximation, 
are given by $\alpha = \Delta N \times$d$\mu_n$/d($N/Z$)/$T$, assuming that the temperature $T$ is the same for
the four reactions. 
Interestingly, the obtained $\alpha$ has less than half the values observed for lithium up to carbon isotope ratios 
in the mass-asymmetric reactions studied by Wada et al.~\cite{wada87} and D\'eak et al.~\cite{deak91}, possibly indicating lower emission temperatures there. % WT
The expected exponential dependence on the binding-energy difference (cf. Eq.~\ref{eq:grandcan}) has been observed 
for the yield ratios of the mirror nuclei t/$^3$He, $^7$Li/$^7$Be, and $^{11}$B/$^{11}$C in Sn+Sn reactions by Xu et al.~\cite{xu00}. % WT

%is only about 20\% smaller than values observed for lithium and boron isotope ratios
%in the mass-asymmetric reactions at quite similar incident energy studied by Wada et al.~\cite{wada87}, possibly indicating lower emission temperatures there.
%The expected exponential dependence on the binding energy (cf. Eq.~\ref{eq:grandcan}) has been observed for the isobar ratios $t/^{3}He$, $^{7}Li/^{7}Be$, and $^{11}B/^{11}C$ in Sn+Sn reactions by Xu et al. \cite{xu00}. % WT

The obtained source properties $T$, $\mu_n$, and $\mu_p$
compare well with values given by the more sophisticated statistical models QSM (Quantum Statistical Model~\cite{hahn88}) and
SMM (Statistical Multifragmentation Model~\cite{SMM}) for sources of low density.
In these models, chemical potentials as derived here from the yield ratios of light fragments and temperatures 
around 6 MeV refer to excited sources 
with density near or below one third of the nuclear saturation density (cf. Figs.~4 and 6 in Ref.~\cite{Botvina02}). 
Freeze-out densities of this magnitude have been deduced from measured correlation functions~\cite{gustaf84,fritz99,verde06} 
and are successfully used in statistical descriptions of multifragmentation data 
(see, e.g., Refs.~\cite{le_fevre04,ogul23,botvina95,raduta07} and references given therein).
%Since the chemical potentials depend only weakly on temperature~\cite{Botvina02}, 
%their comparison with the mean-field results for infinite matter reported by Baran et al.~\cite{Baran02} also supports
%low densities of the emitting medium, and reveals a sensitivity to the stiffness of the nuclear equation of state. 

%{\color{red}{On the whole, we can consider that the grand-canonical approximation gives a fairly accurate description of sideward emission rates of isotopes. 
%Nevertheless, given the deviations from the systematics predicted by this approach, 
%we can also assume that the underlying conditions of fragment formation deviates somewhat from pure thermodynamical equilibrium.}}

Summarizing this section, we may conclude that the grand-canonical approximation gives a sufficiently accurate description of the sideward emission rates, permitting a qualitative understanding of the observed yield ratios spread over two orders of magnitude. For improvements, models like the SMM explicitly treating secondary fragment decays may be appropriate. % WT

%\noindent
%{\color{red} forward angles}

Proceeding now from the chemically equilibrated sideward emissions to the yield ratios obtained at forward angles
for the same centrality classes (Fig.~\ref{fig:fig3}, right panel), we find similar trends and  
somewhat different average values but also a significant difference. As a function of the total $N/Z$, a
characteristic step pattern is observed which is common to all eight cases. 
%and, in particular, deviate from the smooth exponential variation observed 
%at sideward angles (Fig.~\ref{fig:fig3}, right panel). 
To highlight the differences, the exponential fits shown in the right panel of Fig.~\ref{fig:fig3}
are constrained to reproduce the slopes found in the left panel.
Compared to the sideward emissions, 
the variation of ratios measured in different reactions is
larger if the projectile is exchanged and smaller if the target is exchanged, indicating an incomplete 
isospin equilibration between the projectile and target matter for the reaction products at forward angles.

These observations are used to derive an isotopic transparency $\tau$ for each of the eight pairs of nuclides
according to the relation

\begin{equation}
\tau = \frac{\tau_P-\tau_T}{\tau_P+\tau_T}
\label{eq:tau}
\end{equation}

\noindent in which $\tau_P$ and $\tau_T$ measure the relative proportions of projectile-like and target-like contributions, respectively, 
to the isotopic compositions that are observed at forward angles. 
%The definition of these coefficients is derived from the exponential behaviour of the dependence of isotope yields on the total $N/Z$ of the colliding system observed at sideward angles.
The definition of these coefficients relies on the correlation of the measured yield ratios with the total $N/Z$ of the colliding system and on the exponential dependence describing it as observed at sideward angles (Fig.~\ref{fig:fig3}, left panel). % WT
For example, $\tau_P$ is obtained as the logarithm of the ratio of the yield ratios from  
two reactions with different projectiles on the same target,
normalized with the difference $\Delta(N/Z)_P = 0.093$ of the neutron-over-proton ratios of the two projectiles according to 

\begin{equation}
\tau_P = \frac{\ln(R_1/R_2)}{\Delta(N/Z)_P}.
\label{eq:tau_P}
\end{equation}

\begin{figure}[!t]	%Fig. 5
  \begin{center}
    \includegraphics[width=10.5cm]{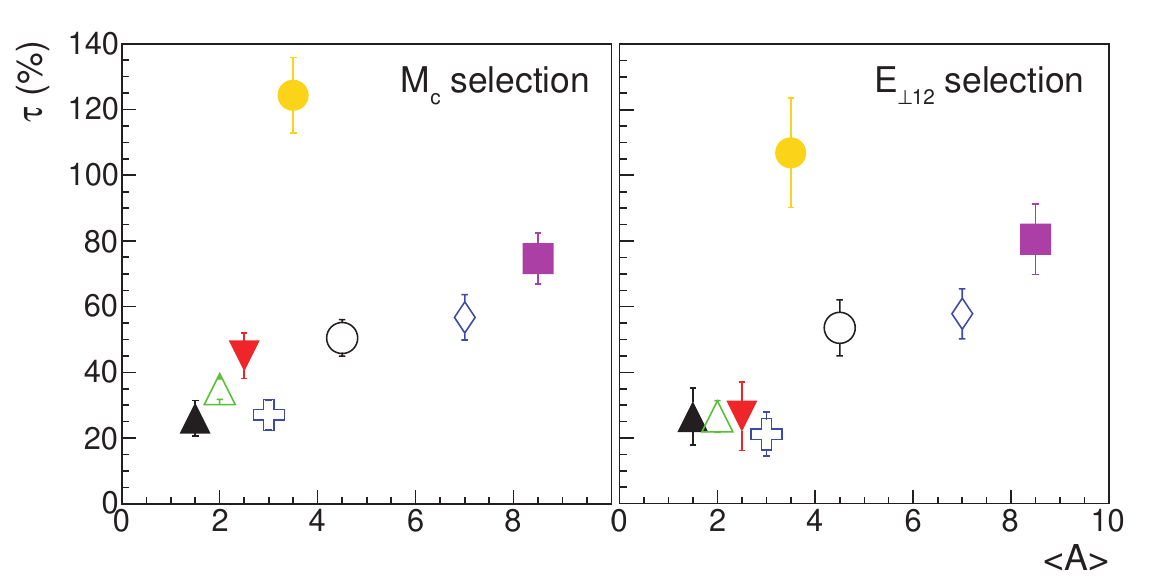} % Fig5.eps for WT
    \caption{Transparency $\tau$ of forward emissions for the studied isotope pairs 
    produced in central collisions of $^{124,129}$Xe+$^{112,124}$Sn at 100~MeV/nucleon selected according 
    to \Mc (left panel) and \Et~(right panel).
    The symbols are chosen as in Figs.~\ref{fig:fig3} and~\ref{fig:fig4} and placed at the mean mass values of the isotope pairs.
    The numerical values are given in Table~\ref{tab:table3}. 
    } 
    \label{fig:fig5}
  \end{center}
\end{figure}

\noindent Here $R_{1}$ and $R_{2}$ denote the two forward yield ratios with $R_{1}$ being that for the case with the more neutron-rich
projectile. This can be done with either one of the two targets, and we use the average value of the two possibilities.
Similarly, $\tau_T$ is obtained from comparing pairs of reaction systems with different targets but the same projectile
and with the difference $\Delta(N/Z)_T = 0.238$ of the two target nuclei as normalization.
In the case of full isospin equilibration, as represented by the exponential fits
at sideward angles, the numerators $\ln(R_1/R_2)$ are equal to ($\alpha\Delta (N_P+N_T))/(Z_P+Z_T)$ and thus
$\tau_P = 0.52 \alpha, \tau_T = 0.48 \alpha$ and $\tau = 0.04$
for the present nearly mass-symmetric reaction systems. Note that $\alpha$ cancels in the expression for $\tau$.
A transparency of 100\%, when the yield ratios do not depend on the target,
causes $\tau_T = 0$ and thus $\tau=1$. A bounce-back would lead to negative values of $\tau$ which, 
however, is neither expected nor observed. The finally obtained values are all positive
(Fig.~\ref{fig:fig5} and Table~\ref{tab:table3}). 

%\noindent
%{\color{red} trends of observed transparencies}

The isotopic transparencies derived with the two methods of choosing the central event group are statistically consistent even
though, for light particles up to $A \le 4$, a preference for slightly smaller values in the case of the \Et~selection
may be present. However, independent of the chosen method, the same systematic trend is observed. 
The transparency is very moderate for hydrogen isotopes and t/$^3$He
($\tau\approx 20-45\%$) whereas transparency values exceeding 50\% are obtained for heavier products.
These results were confirmed by a recent analysis of the same data, selecting central events up to 
4\% of the recorded cross section with a third method and using similar angular cuts~\cite{Genard23}.

One may argue that the incomplete overlap of projectile and target at finite impact parameters can mimic transparencies for geometric reasons. 
In fact, the actual distribution of impact parameters is expected to reach beyond the sharp-cutoff limit $b/b_{max} \le 0.1$ 
corresponding to the selected most central $1\%$ of the cross section. 
Following Ref.~\cite{andronic06}, it is more likely a typical impact parameter rather than a maximum value. 
Even though the incident energy 100 MeV/nucleon is far below the regime at which the participant-spectator model is seriously applicable, 
its predictions may still be useful for estimating the order of magnitude of possible spectator contributions at forward angles. 
According to the formulae presented by Gosset et al.~\cite{gosset77}, the spectator matter appearing at $b/b_{max} \approx 0.1$  
in collisions of approximately equal size nuclei amounts to about $10\%$ which is small with respect to the observed isotopic transparencies.

\begin{table}[!ht]
  \caption{Transparency $\tau$ (in percent) as derived from
    forward emissions ($\theta_{\rm cm}\leq 34^{\circ}$) for the studied isotope pairs 
    produced in central collisions of $^{124,129}$Xe+$^{112,124}$Sn at 100~MeV/nucleon and selected according 
    to \Mc and \Et. Only statistical errors are given. 
  }
  \begin{center}
    \begin{tabular} {l c c c c}
      \hline\hline
      \\
      & d/p & t/d & t/p & t/$^{3}$He \\
      $\tau (\%) $ \Mc sel. & $26\pm5$ & $45\pm7$ & $35\pm3$ & $27\pm5$ \\
      $\tau (\%) $ \Et~sel. & $25\pm9$ & $26\pm10$ & $27\pm5$ & $21\pm7$ \\
      \hline
      \\
      & $^{4}$He/$^{3}$He & $^{6}$He/$^{3}$He & $^{8}$Li/$^{6}$Li & $^{10}$Be/$^{7}$Be \\
      $\tau (\%) $ \Mc sel.& $124\pm12$ & $50\pm6$ & $56\pm7$ & $75\pm8$  \\
      $\tau (\%) $ \Et~sel.& $107\pm17$ & $53\pm9$ & $58\pm8$ & $80\pm11$ \\
      \\
      \hline\hline
    \end{tabular}
  \end{center}
  \label{tab:table3}
\end{table}

%\noindent
%{\color{red} anomalous transparency}

An anomalous transparency exceeding $100\%$ is derived from the $^4$He/$^3$He yield ratio for the selection with $M_c$
and nearly as high for the selection with \Et. 
It is associated with a forward enhancement of the $^4$He emissions contrasting those of $^3$He and 
other light particles with $Z = 1,2$. By using the values for t/$^3$He, the ratios of $^{3,4,6}$He with respect to protons for $N/Z = 1.35$
can be determined from the experimental results listed in Table~\ref{tab:table2} and from the corresponding data for forward emission and
the \Et~selections. The forward-to-sidewards ratio of $^4$He/p is 2.08 (1.52) 
for the case of \Mc (\Et) selections and thus significantly larger than the average forward to sidewards 
ratios $1.13 \pm 0.15$ ($1.03 \pm 0.10$) of the remaining four yield ratios of $Z = 1,2$ particles with respect to protons
(the quoted errors are rms values). 
As hydrogen isotopes exhibit a close to complete momentum relaxation (Fig.~\ref{fig:fig1}), it follows that essentially only the
$^4$He angular distribution is longitudinally extended. Because $\approx 80\%$ of the $Z=2$ yields are $^4$He, this
fact is also evident in Fig.~\ref{fig:fig1}.

The observed $\approx 100\%$ transparency, in addition, indicates that $\alpha$ particles observed at forward angles mostly
originate from the projectile. Their forward enhancement is thus
not unexpected because, similar to neutrons~\cite{pawlowski23}, they are more likely produced in secondary deexcitations 
of heavier products than p, d, t, or $^{3,6}$He particles, thereby carrying part of the longitudinal momenta of the heavy nuclei 
emitting them. A second possible interpretation relates to the participant-spectator picture briefly evoked above.
Alpha clustering of the non-interacting surface parts of low density 
in very central collisions may be another reason 
for the observed high $^4$He/$^3$He transparency~\cite{typel14,tanaka21}. In fact, 
if these parts of the projectile surfaces remain larger in collisions with the smaller $^{112}$Sn target than
in collisions with $^{124}$Sn,
their clusterization may even explain
the inverted target dependence, i.e. larger $^4$He/$^3$He ratios with the neutron-poor $^{112}$Sn target
(Fig.~\ref{fig:fig3}, right panel). The inversion causes the anomalous transparency 
of more than 100\% that is exclusively observed for $^4$He/$^3$He (Fig.~\ref{fig:fig5}).

%\noindent
%{\color{red} comparison with $R_p$}

Finally, comparing the so obtained isotopic transparency with the isotropy ratio $R_p$ (Fig.~\ref{fig:fig1}), we observe very consistent
common trends. The average $R_p \approx 0.7$ for $Z = 1$~and 2 is associated with 
$\tau\approx 30\%$ (Table~\ref{tab:table3}, upper part), both reflecting significant stopping 
and momentum equilibration for the light-particle emission. For the Be isotopes, the largest fragments considered here, the high
transparency exceeding 70\% confirms the
interpretation that the moderate isotropy ratio $R_p \approx 0.4$ is caused by the incomplete stopping of significant amounts of
projectile material, even in the selected very central collisions. The present isotopic study thus supports and complements 
the conclusions reached in the dynamical~\cite{zbiri,le_fevre19} and statistical~\cite{le_fevre04} interpretations of fragmentation 
at the present energy which falls into the transition regime toward the participant-spectator scenario at relativistic energies.

%\noindent
%{\color{red} {\bf Summary}}

In summary, the INDRA multidetector with a solid-angle coverage of nearly $4\pi$ was used in experiments conducted with 
xenon beams of 100 MeV/nucleon
from the SIS18 synchrotron at the GSI laboratory to study isospin equilibration in central collisions in the approximately mass 
symmetric $^{124,129}$Xe+$^{112,124}$Sn reactions.
The $N/Z$ interval covered with the four projectile-target combinations 
extended from 1.27 to 1.43, large enough to observe significant variations of the measured yield ratios
of isotopically resolved reaction products with mass numbers $A \le 10$. Sideward emissions at polar
angles $\theta_{\rm cm} = 90^{\circ} \pm 10^{\circ}$ are found to exhibit isotopic equilibrium. The temperature $T = 6.8$~MeV and the 
chemical potential for neutrons 
$\mu_n = -10.2$ MeV 
%$\mu_n = -10.2 \pm 0.32$ MeV 
derived in grand-canonical approximation suggest a chemical freeze-out at low-density
which does not contradict fragments being dynamically pre-formed at higher densities as suggested in Refs.~\cite{zbiri,le_fevre19,ono19,Genard23} for similar reactions. % WT
%{\color{red} {which does not exclude that fragments are pre-formed at higher densities as suggested in Refs.~\cite{Genard23,zbiri,le_fevre19,ono19} for similar reactions.}} % ALF
%{\color{red} {Interestingly other scenarii (e.g. \cite{Genard23,zbiri,le_fevre19,ono19}) based on higher density dynamical pre-formation of clusters can describe as well their yields.
%One should indeed keep in mind that statistical freeze-out models do not claim that clusters are generated at freeze-out, 
%but rather offer predictions of isotope yields according to thermodynamical conditions when the system reaches freeze-out densities, i.e. when the strong force between clusters becomes negligible, 
%when clusters can escape the medium.}}. % ALF

Similar results are also observed at forward
angles $\theta_{\rm cm} \le 34^{\circ}$, however with the difference that the measured isotope yield ratios indicate a dominance of 
projectile matter. 
The deduced isotopic transparency $\tau$ is found to increase from below $\tau\approx 30\%$ for $Z = 1,2$ particles to a 
significant value exceeding $\tau\approx 70\%$ for $Z=4$ isotopes, the heaviest products included in this study. 
The $^4$He production deviates from the systematic trend, and
possible explanations for the observed anomalously high transparency were presented. The isotopic transparencies are found to be
compatible with the isotropy ratios deduced from measured transverse and longitudinal momentum variances and, very generally,
show that excess momenta in longitudinal direction 
are caused by the incomplete stopping of the projectile, even in the selected highly central collisions at this energy. 
\\

\noindent
{\bf Acknowledgments}

The authors would like to thank the staff of the GSI for providing high-quality $^{124,129}$Xe beams adapted to the requirements 
of the INDRA multidetector. Helpful discussions with A.S. Botvina, P. Napolitani  and S. Typel are gratefully acknowledged. This work was supported 
by the European Community under
contract No. ERBFMGECT950083 and by the French-German Collaboration Agreement 03-45 between IN2P3 - DSM/CEA and GSI.

\end{document}